\documentstyle[preprint,epsf,aps,eqsecnum]{revtex}

\newcommand{\dslash}{\partial\hspace{-.09in}/}

\newcommand{\pslash}{p\hspace{-.07in}/}
\newcommand{\Dslash}{D\hspace{-.11in}/}

\begin{document}

\title{Coherent Neutrino Interactions in a Dense Medium}

\author{Ken Kiers\footnote{Email: kiers@bnl.gov\vspace{-.15in}}}
\address{Department of Physics\\ Brookhaven National Laboratory,
Upton, NY 11973-5000, USA}

\author{Nathan Weiss\footnote{Email: weissn@post.tau.ac.il}}
\address{School of Physics and Astronomy, Tel Aviv University,\\
Tel Aviv, Israel\\Department of Particle Physics, Weizmann Institute,\\
Rehovot, Israel\\ Department of Physics and Astronomy,
University of British Columbia\\Vancouver, B.C., V6T 1Z1  Canada}
 
\maketitle
\begin{abstract}
Motivated by the effect of matter on neutrino oscillations (the MSW
effect)
we study in more detail the propagation of neutrinos in a dense medium.
The dispersion relation for massive neutrinos in a medium is known
to have a minimum at nonzero momentum $p\sim G_F\rho/\sqrt{2}$.
We study in detail the origin and consequences of this dispersion
relation
for both Dirac and Majorana neutrinos both in a toy model with only
neutral currents and a single neutrino flavour and in a realistic 
``Standard Model'' with two neutrino flavours. We find that for a range
of neutrino momenta near the minimum of the dispersion relation, Dirac
neutrinos are trapped by their coherent interactions with the medium.
This effect does not lead to the trapping of Majorana neutrinos.
\end{abstract}
\label{chapter4}

\section{Introduction}
\label{sec:intro}

Motivated by the effect of matter on neutrino oscillations (the 
MSW effect~\cite{msw}),
there have been several works in recent years aimed at understanding
in a more complete way the propagation of one or more flavours of
massive
neutrinos in matter. One of the first papers along these lines was the
paper
of Mannheim in 1987~\cite{mannheim} 
whose main purpose was to derive the MSW
effect from a Field Theoretic starting point.  Mannheim used
second quantization techniques to derive the wave functions and
dispersion
relations of two flavours of both Dirac and Majorana neutrinos
propagating
in a medium in which there was a finite density of electrons. Mannheim
then
analyzed his result in the ultrarelativistic regime and recovered the
standard
MSW results.

Both the work of Mannheim and the work of Nieves~\cite{nieves} 
and of N\"{o}tzold and Raffelt~\cite{notraf}
showed that the entire MSW effect could be reliably analyzed with
a modified Dirac (or Majorana) equation by adding to the frequency 
of the electron neutrino a term
proportional to the density ($\sqrt{2}G_F\rho$ in the standard MSW
scenario).
This term is analogous to a chemical potential term for the electron
neutrino
though it couples only to the chiral left handed part of a possibly
massive neutrino.

In 1991 Panteleone~\cite{pantaleone} 
used such a modified Dirac equation to study the
behaviour of neutrinos in supernova cores. He first analyzed the case of
a
single neutrino flavour without restricting the momenta to be
relativistic. At this point he noticed
a very unusual characteristic
of the neutrino dispersion relation. 
The dispersion relation
had a minimum at a nonzero neutrino momentum. Thus for a range of
neutrino
momentum the neutrino's phase velocity and group velocity  
are in opposite directions!
Panteleone later analyzed the case of two and three flavours of
neutrinos.
He noticed that the neutral currents did not decouple in general 
although
his detailed analysis was carried out only in the high energy regime 
in which they do decouple.

In this paper we provide a synthesis of many of the above results. 
We are particularly
interested in effects at low neutrino momentum and in effects
due to the minimum
of the dispersion relation at nonzero momentum. 
We begin in Section~\ref{sec:4.1} by examining a 
simple model with only a single neutrino flavour
in which the neutrino propagates in a background of
electrons.  We take the neutrino-electron interaction to be
mediated only by neutral current interactions.
This model allows for a careful analysis of the
Field Theory aspects of neutrino propagation in a medium.
We pay special attention to the minimum of the dispersion
relation which occurs at non-zero momentum 
and analyze its effects on neutrino interactions.  We will
find that the minimum energy for Dirac neutrinos in
the medium will generically be {\em less} 
than the rest mass of the neutrino
in the vacuum so that very low energy
neutrinos are effectively
{\em trapped} by the medium.  A similar effect has been 
studied from a different point of view in the work of 
Loeb~\footnote{Loeb studies the problem from 
the point of view of the neutrino's ``index of refraction''
in the medium. A radially varying density leads to a force on the
neutrino which allows it to have bound orbits in the
medium.}\cite{loeb},
although his effect has an entirely different origin.
Our analysis of the trapping of neutrinos will bring
up many interesting questions and puzzles which will
need to be resolved in order to have a complete understanding
of the problem.  We will also examine the case of
Majorana neutrinos and find that in general Majorana neutrinos
are not trapped.

In Sec.~\ref{sec:dispersion} we extend our analysis to 
a more realistic model which has two neutrino flavours
and in which there are both neutral and charged current
interactions.  In this case the dispersion relations are
governed by quartic equations for Dirac neutrinos 
and quadratic equations for Majorana neutrinos.
We will again be most interested in 
examining the form of the dispersion relations in 
the medium and their effects on neutrino propagation.
Again we will find that the Dirac case leads quite
generically to neutrino trapping while the Majorana
case can have no trapping.
We will also make a few remarks regarding the oscillations of neutrinos
in a medium, noting that even for a {\em single} neutrino there
could in principle be oscillations in the probability to detect
the neutrino, due to the differing phase velocities of the helicity
eigenstates.

We conclude in Sec.~\ref{sec:concl} with
a brief discussion of our results and some 
concluding remarks.

\section{A Simple Model with One Neutrino Flavour}
\label{sec:4.1}

Many of the interesting effects which we will discuss,
namely those due to the minimum of the dispersion relation
which occurs at non-zero momentum, occur 
already in a simple model with only a single neutrino flavour.
It is useful, then, to first consider a simplified
model in which a Dirac neutrino propagates in an electron ``gas'' to
which it couples only via the neutral current interaction.  The
case of a Majorana neutrino is somewhat more subtle and will be
considered subsequently.  Our model may be described by the following
Lagrangian 
\begin{eqnarray}
        {\cal L} & = & {\overline \psi}_{\nu}
		\left( i\Dslash^+ - m_{\nu}\right)\psi_{\nu}
		+{\overline \psi}_e
                \left( i\Dslash^- - m_e\right)\psi_e
		 -\frac{1}{4}F^2+\frac{m_Z^2}{2}Z^2
		-\mu_e\psi_e^{\dagger}\psi_e ,
	\label{eq:ltoy}
\end{eqnarray}
where
\begin{eqnarray}
	D_{\mu}^{\pm} & = & \partial_{\mu} \pm i\frac{g}{2\sqrt{2}}
		Z_{\mu}(1-\gamma^5), \\
	F_{\mu\nu} & = & \partial_{\mu}Z_{\nu}-\partial_{\nu}Z_{\mu}.
\end{eqnarray}
The chemical potential term in the Lagrangian is included 
in order to give a non-zero value to the electron density; that is
\equation
	\rho_e = \langle \psi_e^{\dagger}\psi_e \rangle \ne 0.
\endequation

\begin{figure}[tb]
\epsfbox[47 524 505 703]{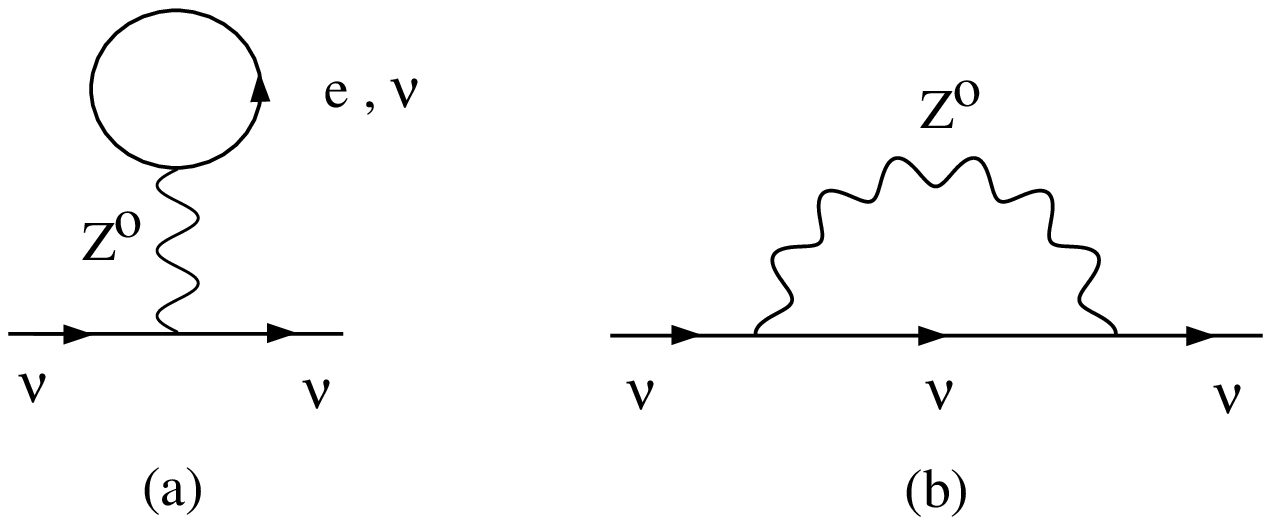}
\caption[One-loop diagrams contributing to the neutrino self-energy in
a simple model]
{One-loop diagrams contributing to the neutrino self-energy in
the model of Sec.~\ref{sec:4.1}.}
\label{fig:oneloop}
\end{figure}

In order to study the propagation of a neutrino in this
medium, we compute the neutrino self-energy.  To one loop there
are three diagrams, shown in Fig.~\ref{fig:oneloop}
\footnote{The electron loop in this diagram is, of course, equal
to the electron density to all orders. The neutrino loop and the $Z_0$
loop lead to
a term proportional to the neutrino density which, as we have pointed
out, is nonzero. This leads to a small correction which can easily be
included but which we shall ignore.}.  All effects
due to the non-zero electron density come from the electron loop
in Fig.~\ref{fig:oneloop}(a) which is easily calculated
and yields
\equation
    \Sigma = - \frac{G}{\sqrt{2}}\rho_e\gamma^0(1-\gamma^5),
\endequation
in which we have defined $G$$=$$g^2/(4\sqrt{2}m_Z^2)$ 
in analogy with the usual 
Fermi coupling constant, $G_F$.  From the self-energy one may
obtain the neutrino propagator in the usual way by summing
a geometric series.  For constant
$\rho_e$ the resulting expression is given in momentum space by
\equation
    G_{\nu}(p) = \frac{1}{\pslash - m_{\nu} -\Sigma}.
\endequation
If $\rho_e$ depends explicitly on $x$ the propagator
may still be formally written in position space as
\equation
    G_{\nu}(x) = \frac{1}{i\dslash - m - \Sigma(x)}.
	\label{proppos}
\endequation
The effective action is then given, to this order, by
\begin{eqnarray}
    S_{\rm eff} &  = & \int d^4x \overline{\psi}_{\nu}\left[
		G_{\nu}^{-1}(x)\right]\psi_{\nu} \\
		& = & \int d^4x \overline{\psi}_{\nu} \left[i\dslash - m
        + \frac{G}{\sqrt{2}}\rho_e\gamma^0(1-\gamma^5)\right]\psi_{\nu}
.
	\label{effaction}
\end{eqnarray}
Variation of the effective action leads finally to 
an effective ``Dirac equation,'' given by
\equation
    \left[i\dslash - m + 
        \alpha\gamma^0(1-\gamma^5)\right]\psi_{\nu} = 0,
    \label{diraceqn}
\endequation    
in which we have defined $\alpha$$=$$G\rho_e/\sqrt{2}$.
For constant electron density, the presence of the ``chiral potential''
in this expression leads to a {\em shift} in the frequency
by $\alpha$, but only for the left-handed (chiral) piece.  This 
shift in the frequency is precisely the 
``index of refraction'' familiar from the MSW effect.
Once we have derived the dispersion relations, it will be clear that
the shift in energy for the neutrino is opposite that for the
anti-neutrino
\footnote{This shift comes from a term in the effective Hamiltonian 
proportional to $\psi^\dagger_\nu\psi_\nu$ which equals the number
density
of neutrinos minus the number density of anti-neutrinos.}.
If the neutrino is ``attracted'' by the medium, then the anti-neutrino
is ``repelled'' by it.

We have noted above that the chemical potential $\mu_e$ in the
Lagrangian~(\ref{eq:ltoy}) gives rise to a non-zero electron
density, $\rho_e$$\equiv$$\langle\psi_e^{\dagger}\psi_e\rangle$.
A similar calculation of 
$\rho_{\nu}$$\equiv$$\langle\psi_{\nu}^{\dagger}\psi_{\nu}\rangle$
for the effective Lagrangian defined by Eq.~(\ref{effaction})
shows that, at least naively, there appears also to be 
a non-zero density of {\em neutrinos} in this medium (provided,
of course, that the effective chemical potential $\alpha$ is larger
than $m$).  It is clear
that this arises due to the potential term proportional to
$\gamma^0$ in the effective Lagrangian, since this term looks
exactly like a chemical potential for the chiral left-handed neutrinos.
How one handles this apparent density of neutrinos
can drastically affect the MSW effect.  Suppose, for example,
that we were to insist that in the sun
$\rho_{\nu}$$\equiv$$\langle\psi_{\nu}^{\dagger}\psi_{\nu}\rangle$$=$$0$.
In order to implement this, we may choose to introduce a ``counter''
chemical potential into the original Lagrangian which
would exactly cancel the neutrino density generated by the interactions
with
the electrons in the medium.  It turns out, however, that this
would change the MSW result quite dramatically.  In fact, if
our theory had a ``vector'' instead of a ``chiral'' potential, we
would kill the entire MSW effect by doing this.
As we shall discuss in more detail below, the correct thing to do is 
to accept the fact that in equilibrium
there is a non-zero density of neutrinos of very low momentum
due to their attractive interaction with the medium~\cite{smirnov}.

Further insight into the physics of our model can be
gained by examining the equations of motion following from
the Lagrangian in Eq.~(\ref{eq:ltoy}).  Varying the Lagrangian
with respect to the $Z^{\mu}$ field leads to
\equation
	\partial_{\nu}F^{\nu\mu} + m_Z^2Z^{\mu} = - \frac{g}{2\sqrt{2}} 
			J^{\mu},
	\label{zmotion}
\endequation
where
\equation
	J^{\mu} =\overline{\psi}_e\gamma^{\mu}(1-\gamma^5)\psi_e 
		-\overline{\psi}_{\nu}\gamma^{\mu}(1-\gamma^5)\psi_{\nu}.
\endequation
If $\rho_e$$=$$\langle\psi_e^{\dagger}\psi_e\rangle$ is constant,
then (\ref{zmotion}) leads to
\equation
	\langle Z^0\rangle = - \frac{g\rho_e}{2\sqrt{2}m_Z^2},
	\label{zexp}
\endequation
that is, the $Z^0$ field has gained a vacuum expectation
value.  The equation of motion for the neutrino field
is then given by
\equation
	\left[ i\dslash -m - \frac{g}{2\sqrt{2}}
		\langle Z^0\rangle\gamma^0(1-\gamma^5)
		\right]\psi_{\nu} = 0,
\endequation
which is equivalent to the effective Dirac equation of (\ref{diraceqn})
once the value for the $Z^0$ expectation value, Eq.~(\ref{zexp}), 
is inserted.  From this point of view, then, the left-handed neutrino
sees a 
mean (coherent) ``scalar potential,'' $\langle Z^0\rangle$.  From
the point of view of the field theoretic calculation above,
it is clear why the $Z^0$ field has developed a vacuum expectation
value.  The one-loop diagram corresponding to $\langle Z^0\rangle$
is simply the electron loop, corresponding to $\rho_e$, with a $Z$
propagator
attached.  This coherent $Z^0$ field is analogous to the 
electric field which surrounds
a static charge distribution and is due to the net
weak charge of the medium.

\subsection{Solution to the Dirac Equation}
\label{sec:soldirac}

\begin{figure}[tb]
\epsfbox[61 410 495 703]{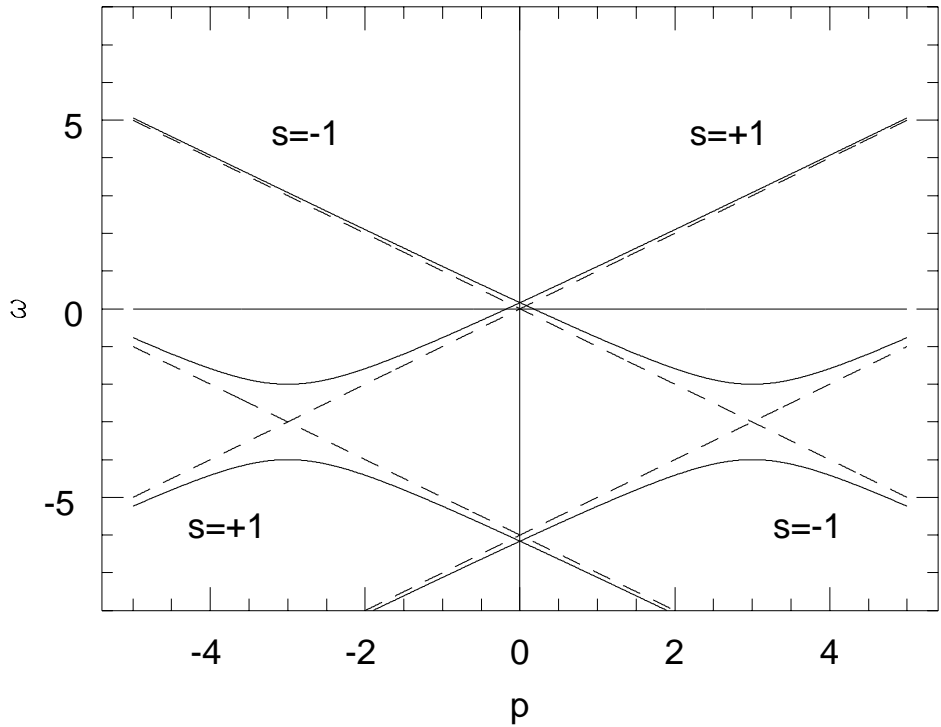}
\caption[Dispersion relations for a neutrino in an electron ``gas'']
{Dispersion relations for the model considered in
Sec.~\ref{sec:soldirac},
with $\alpha$$=$$3$ and $m$$=$$0, 1$, in arbitrary units.
The dashed and solid curves correspond, respectively, to the $m$$=$$0$
and $m$$\ne$$0$ cases. Note how the curves in the massive case 
result from 
the level repulsion at the level crossings of the massless case.}
\label{fig:disp1}
\end{figure}

In order to study the propagation of neutrinos over macroscopic
distances, it suffices to study the effective Dirac equation
given in Eq.~(\ref{diraceqn}).  The propagator defined in 
Eq.~(\ref{proppos}) contains this information, but it also encodes
the off-shell behaviour of the neutrino.  
For a medium with constant density, it is straightforward to solve the
Dirac equation in momentum space by employing the chiral representation,
so that
\equation
	\psi =\left(\begin{array}{c} \chi_L \\ \chi_R \end{array}\right),
\endequation
in which the upper and lower components correspond to the left and
right chiral projections, respectively.
In this representation, the Dirac equation becomes
\equation
	\left(\begin{array}{cc}
		-m & \omega -\vec{\sigma}\cdot \vec{p} \\
		\omega +2\alpha +\vec{\sigma}\cdot \vec{p} & -m
		\end{array}\right)
	\left(\begin{array}{c} \chi_L \\ \chi_R \end{array}\right) = 0.
\endequation
where $\alpha$$=$$G\rho_e/\sqrt{2}$.
Without loss of generality we may choose $\vec{p}$$=$$p\hat{z}$,
so that $\chi_{L,R}$ (and hence also $\psi$) 
may be chosen to be eigenstates of 
$\sigma_3$, the spin projection in the $z$ direction.  That is,
\equation
	\sigma_3 \chi_{L,R} = s\chi_{L,R},
\endequation
where $s$$=$$\pm 1$.  Solving for the energy yields four solutions
\equation
	\omega = -\alpha \pm \sqrt{(p+\alpha s)^2+m^2}.
	\label{energies}
\endequation
These dispersion relations are plotted in Fig.~\ref{fig:disp1}
both for $m$$=$$0$ (dashed curves)  and $m$$\ne$$0$
(solid curves.)  Several key features of these plots
should be noted.  First of all, the ``negative energy'' states
are, in this case, those which are unbounded from below as 
the momentum is increased.  In the second quantized theory the
correct energy of such a state is just the negative of
its energy eigenvalue.  We also note that when 
$m$$=$$0$ there are ``level crossings.'' 
These are avoided for $m$$\ne$$0$ by level repulsion
due to the mixing of the levels.

The most noteworthy feature of these dispersion relations
(discussed previously by Pantaleone) is the fact that
the  minima of the dispersion relations occur at
non-zero values of the momentum, $p$$=$$\pm\alpha$, instead
of at the origin.
One interesting consequence of this fact is that
the neutrino can have a vanishing group
velocity at non-zero momenta.  Furthermore, for $|p|$$<$$|\alpha|$,
the neutrino's group velocity, $d\omega/dp$, is in a direction opposite
to its
momentum!  This will play an important role in understanding the
reflection of 
neutrinos at the boundary of the medium.  Another interesting feature 
of these curves
is that the minimum energy $\omega_{\rm min}$$=$$-\alpha+m$ is
{\em less than} the neutrino mass.  Thus it is possible to
produce a neutrino in the medium which has 
$\omega$$<$$m$.  Such a neutrino will not
have enough energy to survive in the vacuum and will thus be
{\em trapped} by the medium.  We shall examine these
peculiar features of the neutrino dispersion relations
in detail below.

Finally, we note that
for high momentum, the solution corresponding to negative helicity has
energy
\equation
	\omega \approx p + \frac{m^2}{2p} -2\alpha ,
\endequation
which is just the usual MSW result.  By way of contrast, the positive
helicity
solution approaches its vacuum value of 
$\omega$$\approx$$p+m^2/2p$.  This illustrates the spin-dependence of
the interaction.  For high momentum, the left-handed (chiral) states
are nearly equivalent to the negative helicity eigenstates, so the
potential (which is left-handed) affects only the negative, and not
the positive, helicity eigenstates.

\subsection{Neutrino Trapping}

\begin{figure}[tbp]
\epsfbox[80 90 513 704]{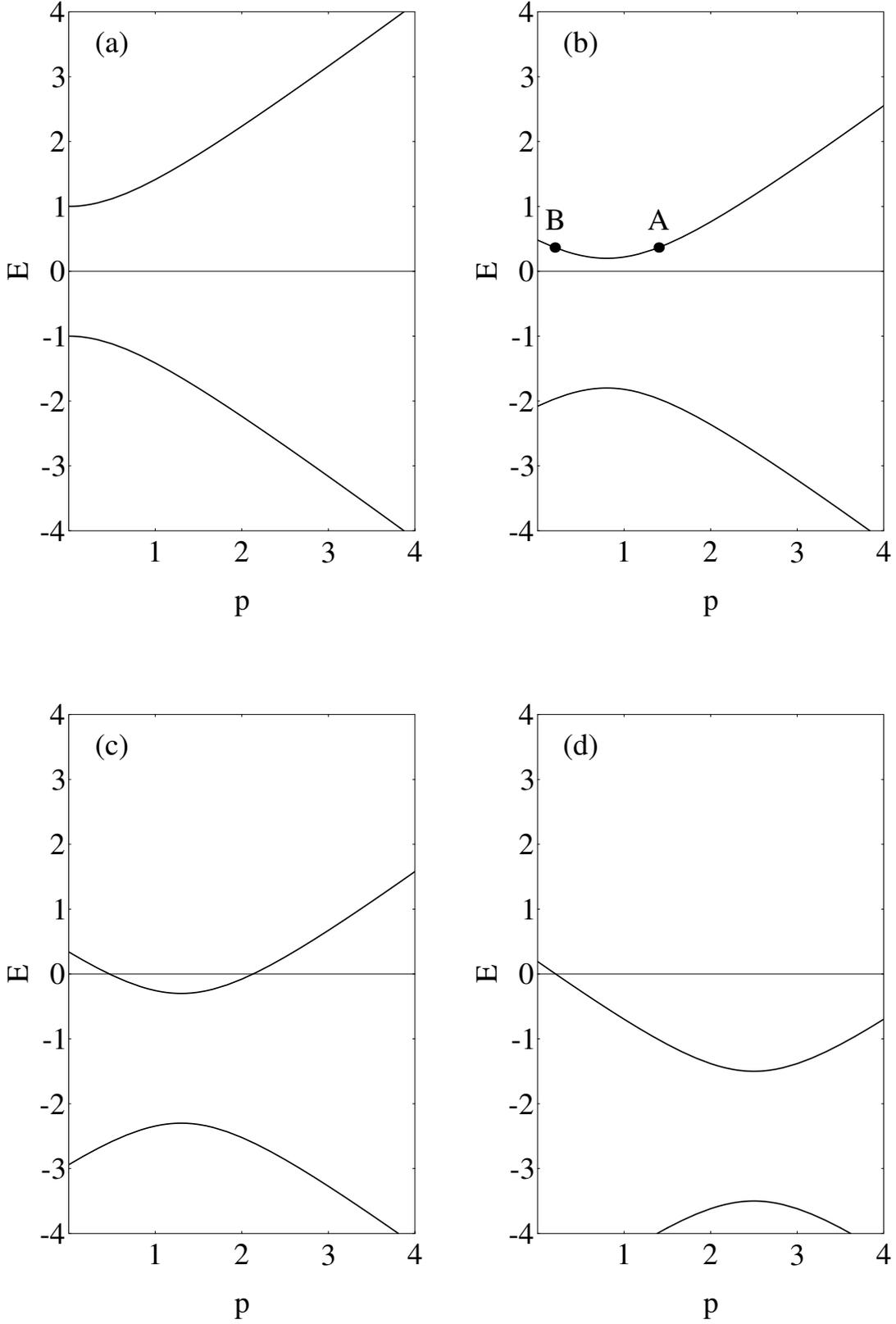}
\caption[Caption]
{Dispersion relations for a neutrino (a) outside and (b)-(d) inside
the medium in the simple model of Sec.~\ref{sec:soldirac}, with
$m$$=$$1$ and $\alpha$$=$$0$, $0.8$, $1.3$ and $2.5$ for 
(a), (b), (c) and (d), respectively.}
\label{fig:levels}
\end{figure}

Let us now consider a low energy neutrino which is produced inside 
the medium and which then tries to escape to the vacuum.  We take the
medium to have~\footnote{Note that for $\alpha$$<$$0$ the medium
would trap anti-neutrinos instead of neutrinos.} 
$\alpha$$>$$0$.  The dispersion relations for this
neutrino both inside and outside the medium are illustrated in 
Fig.~\ref{fig:levels}. In this figure the neutrino energy as a function
of
its momentum is plotted for zero density (outside the medium) 
(Fig.~\ref{fig:levels}(a)), low density (Fig.~\ref{fig:levels}(b)),
intermediate density (Fig.~\ref{fig:levels}(c)) and high density
(Fig.~\ref{fig:levels}(d)). The negative energy solutions to the
Dirac Equation are also plotted, for reasons which will 
become apparent momentarily. 

Let us begin by looking at the case of low electron density,
$0$$<$$\alpha$$<$$m$. This region is characterized by the fact that the
minimum
of the neutrino dispersion relation lies above the vacuum energy
($E$$=$$0$)
but below the neutrino mass ($E$$=$$m$).
In this case a neutrino which is produced inside the medium
with momentum $p$ such that $E(p)$$>$$m$ will escape from the medium
(although
there will also be some amplitude for reflection). 
If, on the other hand, its momentum
$p$ is close to $\alpha$ so that $E(p)$$<$$m$ (as is shown, for example,
by
the point A in Fig.~\ref{fig:levels}(b)), then this 
neutrino is trapped in the medium
since there is no Energy level corresponding to its energy if it escapes
into the vacuum (Fig.~\ref{fig:levels}(a)). 
The way this total reflection is realized in practice is
fascinating.  Suppose the neutrino is incident normal to the interface
of the
medium and the vacuum.  We might then
be concerned that the neutrino has to flip
its spin in order to find an energy level with negative momentum {\bf
even
in the medium} (see Fig.~\ref{fig:disp1}.)  It is straightforward
to show, however, that the neutrino's spin must be conserved
in such a reaction. The resolution of this apparent problem is that 
the reflected
neutrino indeed has a {\bf positive} momentum as shown by the point B  
in Fig.~\ref{fig:levels}(b). 
The reason this corresponds to a reflected wave is that the neutrino's
{\bf group} velocity is negative in this region so that the neutrino
travels
back into the medium.  It is straightforward to extend these arguments
to the case of non-normal incidence~\cite{thesis}.

The case of intermediate electron density, $m$$<$$\alpha$$<$$2m$,
is shown in Fig.~\ref{fig:levels}(c) and is characterized by the fact 
that the
minimum neutrino energy in the medium is less than zero but greater than
the largest negative energy state outside the medium. In this case the
lowest
energy of the system occurs when all the neutrino states below $E$$=$$0$
are filled so that the mean neutrino density $\rho_\nu$ is nonzero.
This is consistent with the field theoretic calculation of this density
which
was discussed previously in this paper. In this case the neutrinos with 
$m$$>$$E$$>$$0$ would be trapped as in the previous case but no
neutrinos with
$E$$<$$0$ could be produced in the medium since all corresponding levels
are
full. One could imagine a {\it nonequilibrium} (higher energy) 
situation in which these levels are not full. In this case neutrinos 
with energy $E$$<$$0$ (due to their attractive interaction with the
medium)
could be produced (by some external reaction) and they
would also be trapped. Whether such a situation arises in any practical
case depends on the dynamics of the formation of this region.

Finally we turn to the case of high electron density for which
$\alpha$$>$$2m$.
This situation is shown in Fig.~\ref{fig:levels}(d) and is characterized
by the
fact that the minimum of the neutrino dispersion relation inside the
medium
has a lower energy than the maximum energy of the negative energy Dirac
states
in the vacuum. In this case the potential difference between the vacuum
and the
medium is strong enough to induce pair production of 
neutrino--antineutrino pairs
at the interface. If the neutrino levels below $E$$=$$-m$ are not filled 
initially then this pair production would eventually lead to the filling
of
these
levels. If a neutrino with energy below $-m$ is produced by another
mechanism
during this time, it would be trapped by the medium. \footnote{In this
case the
reflection amplitude would also have a contribution from pair production
which would allow a spin flip at the boundary~\cite{hansen}. 
This contribution will be the 
dominant contribution in the case of massless neutrinos.}

In all the cases discussed above it is simple to write down the
criterion
which must be met in
order that a neutrino be trapped.   
The dispersion relation for a negative spin (relative to $\hat{z}$) 
neutrino is given by
\equation
	\omega = -\alpha + \sqrt{(p-\alpha)^2+m^2},
\endequation
so that the condition for trapping is
\equation
	-\alpha + \sqrt{(p-\alpha)^2+m^2} < m
\endequation
or
\equation
	p < p_{\rm trap} \equiv \alpha + \sqrt{\alpha^2 + 2\alpha m}.
	\label{ptrapdef}
\endequation
Thus, neutrinos produced in this medium with momentum $p$$<$$p_{\rm
trap}$
will not have enough energy to survive in the vacuum and will
be trapped.   

Before proceeding it is useful to get some idea
of the overall magnitude of the effect of neutrino trapping.
Setting $G$$\approx$$G_F$ and $m$$\approx$$10^{-3}$eV (which is 
a mass relevant for the MSW-resolution of the solar neutrino problem),
we find that
$p_{\rm trap}$$\sim$$10^{-8}$eV in the sun (for
which $\alpha$$\sim$$10^{-12}$eV) and 
$p_{\rm trap}$$\sim$$100$eV in a supernova (for
which $\alpha$$\sim$$100$eV.)  The phenomenon of trapping
is quite remarkable when we consider that the 
mean free path (which increases with decreasing momentum)
of a neutrino with $p$$\sim$$10^{-8}$eV in
the sun is on the order of $10^{20}$ {\em solar radii}.  Such a neutrino
would thus have no chance of being ``incoherently'' trapped in the sun
(say by back-scattering from nuclei), but would still be trapped by the
coherent process which we are discussing.  We note furthermore that
in the case of incoherent trapping
the scattering cross section is typically dependent on the mass of 
the target particle, whereas for coherent trapping this is not the
case.  Loeb has discussed a similar effect (albeit of different origin)
for neutron stars and
has estimated that in general this effect will add an extra 30 kg to the
mass
of the star~\cite{loeb}.
 
\subsection{The Majorana Case}
\label{sec:maj1}

In the case of Majorana neutrinos, there is only a single
(left-handed) field, $\chi_L$.
The effective Lagrangian is given by
\equation
	{\cal L}_{\rm eff} = \overline{\psi}_M
		\left[ \frac{1}{2}(i\dslash - m) +\alpha\gamma^0
		(1-\gamma^5)\right]\psi_M,
\endequation
where
\equation
	\psi_M =\left(\begin{array}{c}
		\chi_L \\ -i\sigma^2\chi_L^* \end{array}\right)
\endequation
in the chiral representation.  In general one needs to be somewhat
careful when dealing with Majorana fermions.  For example, even
at the classical level, the fields
need to be taken as Grassman-valued, or else the mass term disappears.
The dispersion relations in this case 
can be obtained by solving the equations of motion, as was
first done by Mannheim~\cite{mannheim}.  The reader is referred
to Mannheim's paper for details of the calculation (see also 
Ref.~\cite{mann2}.)
The resulting expression for the negative helicity neutrino is given by
\equation
	\omega = \pm\sqrt{\left(|\vec{p}|-2\alpha\right)^2+m^2}.
\endequation
In this case the energy has a minimum value,
$\omega$$=$$m$, which occurs at $|\vec{p}|$$=$$2\alpha$.
In fact, the dispersion relation in matter is identical to that in
vacuum except for a lateral shift to the right.  This implies in
particular that, in contradistinction to the Dirac case,
a neutrino cannot have an energy less than $m$ and 
there is thus {\em no trapping} of Majorana neutrinos
in the medium.

\section{Dispersion Relations for Two Neutrino Flavours}
\label{sec:dispersion}

We turn now to consider a more realistic scenario in which 
there are two neutrino flavours and in which there are both
neutral current and charged current couplings to the medium.
We have in mind, of course, the Standard Electroweak Model
with massive neutrinos.
We shall first derive the quartic equation governing the
dispersion relations in the Dirac case and examine the
solutions in some representative cases.  We shall see again
that in this case there is neutrino trapping.  We then 
examine the Majorana case, in which the dispersion relations
are quadratic and are thus readily analyzed but in which both
trapping and pair production at the surface of the sun are absent. 
We shall then describe briefly
an alternate model which
has been studied in the literature which has {\em no} chiral coupling
but still yields a minimum in the dispersion relation at
non-zero momentum.  Finally, we comment on neutrino oscillations
in these models.

\subsection{Dirac Case}

\begin{figure}[tb]
\epsfbox[47 524 505 703]{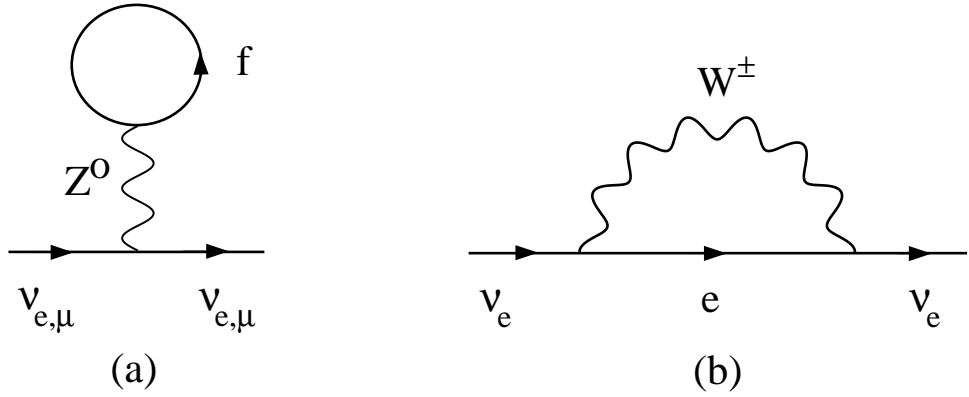}
\caption[One-loop diagrams contributing to the neutrino self-energy in
a realistic model]
{Density-dependent one-loop diagrams contributing to the neutrino
self-energy in a realistic model with both neutral-current and
charged-current
couplings. In (a) the ``f'' stands for the contributions due to all
fermions in the medium which have neutral current couplings.  In (b) we
have assumed that the medium contains electrons and positrons, but
no other charged leptons.}
\label{fig:oneloop2}
\end{figure}

We begin with the Dirac Equation in the mass basis
for a pair of massive Dirac neutrinos with
both neutral and charged current coupling to a medium:
\equation
	\left\{\pslash -M +\left(\beta-\alpha Q\right) \gamma^0\left(
	1-\gamma_5\right)\right\}\psi=0
	\label{deqn2neut}
\endequation
where $M$ is the {\bf diagonal} 
$2\times 2$ mass matrix, $\beta\propto \rho G_F$
is the contribution of the neutral current which couples only to the
left
handed neutrinos and $\alpha\propto \rho_e G_F$ represents the charged
current
contribution which couples only to $\nu_e$.  This coupling is assured by
the 
mixing matrix
\equation
	Q=\left(\matrix{\cos^2(\theta)& \sin(\theta)\cos(\theta)\cr
	\sin(\theta)\cos(\theta)&\sin^2(\theta)}\right) .
	\label{qdef}
\endequation
$\alpha$ and $\beta$ may be calculated by computing the one loop
contributions to the neutrino 
self-energy in the background medium.  The Feynman diagrams
corresponding
to these processes are shown in 
Fig.~\ref{fig:oneloop2} and yield~\cite{nieves,notraf,palpham}
\equation
        \alpha = \frac{G_F}{\sqrt{2}}\rho_e
\endequation
and
\equation
        \beta = -\frac{G_F}{\sqrt{2}}
                \sum_{f}(T_3^{(f)} -2Q^{(f)}{\rm sin}^2\theta_W )\rho_f
	\label{betadef1}
\endequation
in which the sum in (\ref{betadef1}) runs over all fermions in the 
medium, $T_3^{(f)}$ is the third component of the fermion's 
weak isospin and $Q^{(f)}$ is its charge.  If there are appreciable
densities of anti-particles in the medium, then $\rho_f$ needs
to be replaced by $\rho_f$$-$$\rho_{\overline{f}}$ in these expressions.

In the Chiral representation we write
\equation
	\psi = \left( \matrix{\chi_L \cr \chi_R}\right)
\endequation
and the Dirac Equation becomes
\equation
	\left( \omega-\vec \sigma\cdot \vec p\right) \chi_R = M \chi_L ,
	\;\;\;\;\;
	\left\{ \left( \omega+\vec \sigma\cdot \vec p\right) +
	2\left(\beta-\alpha Q\right) \right\}\chi_L = M \chi_R .
\endequation
For simplicity we may assume the momentum to be in the $\hat{z}$
direction
in which case the solutions to the Dirac Equation will be eigenstates
of $\sigma_3$:
\equation
	\chi_L=\left( \matrix{L_+ \cr 0 }\right),\;\;
	\chi_R=\left( \matrix{R_+ \cr 0 }\right);
	~~~~~~{\rm or}~~~~~
	\chi_L=\left( \matrix{0 \cr L_- }\right),\;\;
	\chi_R=\left( \matrix{0 \cr R_- }\right)
\endequation
leading to the equations:
\equation
	\left\{\omega^2-p^2+2(\omega\mp p)\left(\beta-\alpha Q\right) -M^2
	\right\}L_\pm =0 ,
	\label{lplusmin}
\endequation
\equation
	R_{\pm}={1\over{\left(\omega\mp p\right)}}ML_{\pm} .
\endequation

To find the energy eigenvalues we rewrite Eq.~(\ref{lplusmin}) as:
\equation
	\left\{\omega^2-p^2-\mu^2 +2(\omega\mp p)\beta -
	2\alpha(\omega\mp p)N_\mp \right\}L_{\pm}=0
\endequation
where $\mu^2=\langle{m^2}\rangle=(m_1^2+m_2^2)/2$ is the mean squared
mass and 
\equation
	N_\mp = \left(\matrix{\cos^2(\theta)-\xi_\mp& \sin(\theta)\cos(\theta)
	\cr \sin(\theta)\cos(\theta) & \sin^2(\theta)+\xi_\mp}\right)
\endequation
with
\equation
	\xi_\mp = {{ \Delta^2}\over{4\alpha(\omega\mp p)}}
\endequation
and $\Delta^2=m_2^2-m_1^2$.
It thus remains only to find the eigenvalues of $N_\mp$.

The eigenvalues of $N_\mp$ are:
\begin{eqnarray}
	\lambda_1^\mp & = & {1\over 2}\left(1+\sqrt{
	1-4\xi_\mp\left(\cos(2\theta)-\xi_\mp\right)}\right) \\
	\lambda_2^\mp & = & {1\over 2}\left(1-\sqrt{
	1-4\xi_\mp\left(\cos(2\theta)-\xi_\mp\right)}\right) .
\end{eqnarray}
Plugging these back into the Dirac Equation leads to the following
quartic equation for the energy eigenvalues:
\equation
	\left[\omega^2-p^2-\mu^2+(2\beta-\alpha)(\omega-sp)\right]^2
	=\alpha^2(\omega-sp)^2-\alpha\Delta^2\cos(2\theta)(\omega-sp)
	+\frac{1}{4}\Delta^4 ,
	\label{quartic}
\endequation
in which $s$$=$$\pm 1$ is the eigenvalue of $\sigma_3$, the spin
projection in the $+z$ direction.  In special cases 
this expression reduces to
those found in the papers of Mannheim (in which the neutral
current contribution has been left out) and Pantaleone (in which one
of the masses has been set to zero.)

Equation~(\ref{quartic}) is the main quartic equation whose eight
solutions
(for $s$$=$$\pm 1$) lead to the eight dispersion relations (four
positive
energy and four negative energy) in the medium.
It may not at first be obvious that all eight of the 
solutions $\omega$ of (\ref{quartic}) corresponding to a fixed value of
the 
momentum are real.  This is, however, the case, which may be seen as
follows.  (The proof is equally straightforward for
any number of flavours, so we will do it immediately in the general
case.)
In the general case the Dirac Equation
in the mass basis becomes
\equation
        \left\{\pslash -M +\left(\beta-
	\alpha{\cal U}^{\dagger}N_e{\cal U}\right) \gamma^0\left(
        1-\gamma_5\right)\right\}\psi=0
\endequation
in which ${\cal U}$ is the mixing matrix in flavour space
and $N_e$$=$${\rm diag}(1,0,\ldots,0)$ is also a matrix in 
flavour space.  Pre-multiplying this
expression by $\gamma^0$ leads to the eigenvalue equation
\equation
	{\cal N}\psi = \omega \psi,
	\label{hermeig}
\endequation
where
\equation
	{\cal N} = \gamma^0\vec{\gamma}\cdot\vec{p} +\gamma^0M
	-\left(\beta -\alpha{\cal U}^{\dagger}N_e{\cal U}\right)(1-\gamma^5).
\endequation
Since ${\cal N}$ is hermitian for real $\vec{p}$, 
the eigenvalues of Eq.~(\ref{hermeig}) are guaranteed to be real.
This completes the proof.

It is best to analyze the expression governing the dispersion
relations, Eq.~(\ref{quartic}),
by first considering some special
cases.  The very simplest case is when $m_1$$=$$m_2$$=$$0$, which yields
\begin{eqnarray}
        \omega & = & p + (\beta-\alpha)(s-1) \\
        \omega & = & -p -(\beta-\alpha)(s+1)
\end{eqnarray}
for the electron neutrinos and
\begin{eqnarray}
	\omega & = & p + \beta(s-1) \\
	\omega & = & -p -\beta(s+1) 
\end{eqnarray}
for the muon neutrinos.  These expressions are easy to understand.
Four of the dispersion relations are unchanged from their
values in the vacuum, since positive helicity neutrinos are
also right-handed (chiral) and are thus unaffected by the 
left-handed Standard Model interactions.
The remaining four dispersion relations
are displaced vertically from their vacuum
values by amounts proportional to their couplings to the medium.
Note that only the dispersion relation corresponding to
$\nu_e$ is affected by the charged current contribution, $\alpha$.

Another simple case occurs when the coupling $\theta$ is set
to zero.  In this case the dispersion relations for $\nu_e$ and
$\nu_{\mu}$ decouple, as one would expect, and we find
\equation
	\omega = -(\beta-\alpha)\pm\sqrt{(p+s(\beta-\alpha))^2+m_1^2}
	\label{unc1}
\endequation
and
\equation
        \omega = -\beta\pm\sqrt{(p+s\beta)^2+m_2^2}
	\label{unc2}
\endequation
for the electron and muon neutrinos, respectively.  These expressions 
are in exact agreement with what we found in the single-neutrino case
in Sec.~\ref{sec:soldirac}.  Once again the dispersion relations
corresponding to the massless case undergo ``level repulsion'' when
a finite mass is added.

Since the equations governing the dispersion relations in
the two-flavour Dirac case are quartic it is difficult,
in general, to obtain analytic
expressions for these dispersion relations.  
Of course quartic equations {\em are} analytically
solvable and we know that in our case all the 
solutions are real.  In general, however, no practical
insight can be gained by examining the analytic expressions of
these solutions.  One approach which is  helpful in understanding
the dispersion relations if the
coupling $\theta$ is not too large is to use a graphical
approach.  One begins by looking at the solutions when $\theta$$=$$0$
in which case the two neutrino flavours decouple and we can use the
solutions derived in the previous section. 
Thus, for example, in Fig.~\ref{fig:quartic}(a) the dotted curves
represent the solutions for $\theta$$=$$0$. Note that the dispersion
relations for the two flavours of neutrinos cross at some points.
When $\theta$ is ``turned on'' we expect that these levels will repel
and will lead to a curve similar to the solid curve in that figure.
The solid curve is, in fact, the solution to the quartic equation when
$\theta$$=$$0.2$. This graphical method is reasonably accurate when
$\theta$
is small but can be used as a guide even for larger values of $\theta$.

One interesting feature in the two-neutrino case is that
the heavier neutrino, which would have been trapped in 
certain cases if the neutrinos were decoupled, can now ``leak out'' due
to its coupling to the lighter mass eigenstate.  That is, only states
with energy less than the mass of the {\em lightest} mass eigenstate are
strictly ``trapped'' now.

It is also possible to derive approximate solutions of the
quartic equations if the neutrinos are relativistic.  In that 
case approximate solutions are given by
\begin{eqnarray}
	\omega & \simeq & p -(2\beta -\alpha)+\frac{\mu^2}{2p}
	\pm\frac{1}{4p}\left[\left(4\alpha p-\Delta^2\cos(2\theta)\right)^2
	+\Delta^4\sin^2(2\theta)\right]^{1/2}, \label{exp1}\\
        \omega & \simeq & -p -(2\beta -\alpha)-\frac{\mu^2}{2p}
        \mp\frac{1}{4p}\left[\left(4\alpha
p+\Delta^2\cos(2\theta)\right)^2
        +\Delta^4\sin^2(2\theta)\right]^{1/2},\label{exp2} \\
        \omega & \simeq & \pm\left(p+\frac{m_{1,2}^2}{2p}\right),
	\label{exp3}
\end{eqnarray}
where the corrections to the above expressions go like
$\alpha\mu^2/p^2$, $\beta\mu^2/p^2$ and $\mu^4/p^3$.
Of these expressions, (\ref{exp1}) gives the energy of the
negative-helicity
particle eigenstates and (\ref{exp2}) gives the energy of the
positive-helicity
{\em anti}-particle eigenstates.  These are in agreement with the
usual result and show that the neutral current contribution
``factorizes''
in the relativistic limit; that is, the difference between
the two negative-helicity particle energies is independent of $\beta$.
Furthermore, it is clear that, if $\alpha$$>$$0$ (which occurs if the
background contains more electrons than positrons) then
a resonance can occur when $4\alpha p$$=$$\Delta^2\cos(2\theta)$.
This is the well-known MSW resonance.
The remaining dispersion relations, given in Eq.~(\ref{exp3}),
are unchanged from their vacuum values (since the
potential is left-handed) and are related to the
positive-helicity neutrinos and negative-helicity anti-neutrinos.

\subsection{Majorana Case}
\label{sec:maj2}

The case of Majorana neutrinos is interesting for two 
reasons.  First of all, it is the favoured realistic scenario
in models which have massive neutrinos, for example in models
which employ the ``see-saw'' mechanism.  Secondly, it turns out
that the equations governing the dispersion relations 
are {\em quadratic} rather than {\em quartic}, which means that
in principle they should be easier to analyze.

The calculation proceeds in a manner similar to 
that followed in Sec.~\ref{sec:maj1}, the only complication
being the additional mixing in flavour space.  We omit the
details and simply present the result.  The negative-helicity dispersion
relations in this case are determined by the equation
\equation
	\left(\omega^2-p^2-\Delta_+^2(p)\right)
	\left(\omega^2-p^2-\Delta_-^2(p)\right) = 0,
\endequation
where
\begin{eqnarray}
	\Delta_{\pm}^2(p) & = & \frac{1}{2}\left(m_1^2+m_2^2+
		4p(\alpha-2\beta)+4\alpha(\alpha-2\beta)
		+8\beta^2\right) \nonumber \\
	& & \;\;\pm\frac{1}{2}
		\left\{\left[(m_2^2-m_1^2)\cos(2\theta)-4\alpha p
		-4\alpha(\alpha-2\beta)\right]^2 \right. \nonumber \\
	& & \;\;\;\;
		+(m_2-m_1)^2\left.\left[(m_1+m_2)^2+4\alpha^2
		\right]
		\sin^2(2\theta)\right\}^{1/2}.
\end{eqnarray}
Thus the four solutions are
\begin{eqnarray}
	\omega & = & \pm \sqrt{p^2+\Delta_{+}^2(p)}, \\
	\omega & = & \pm \sqrt{p^2+\Delta_{-}^2(p)}.
\end{eqnarray}
These again reduce to Mannheim's result if we set 
$\beta$$=$$0$~\cite{mannheim}.
It is interesting to note that these solutions are not
functions only of $m_1^2+m_2^2$ and $m_2^2-m_1^2$, as is
the case in the relativistic regime.  

\begin{figure}[tbp]
\epsfbox[50 197 457 698]{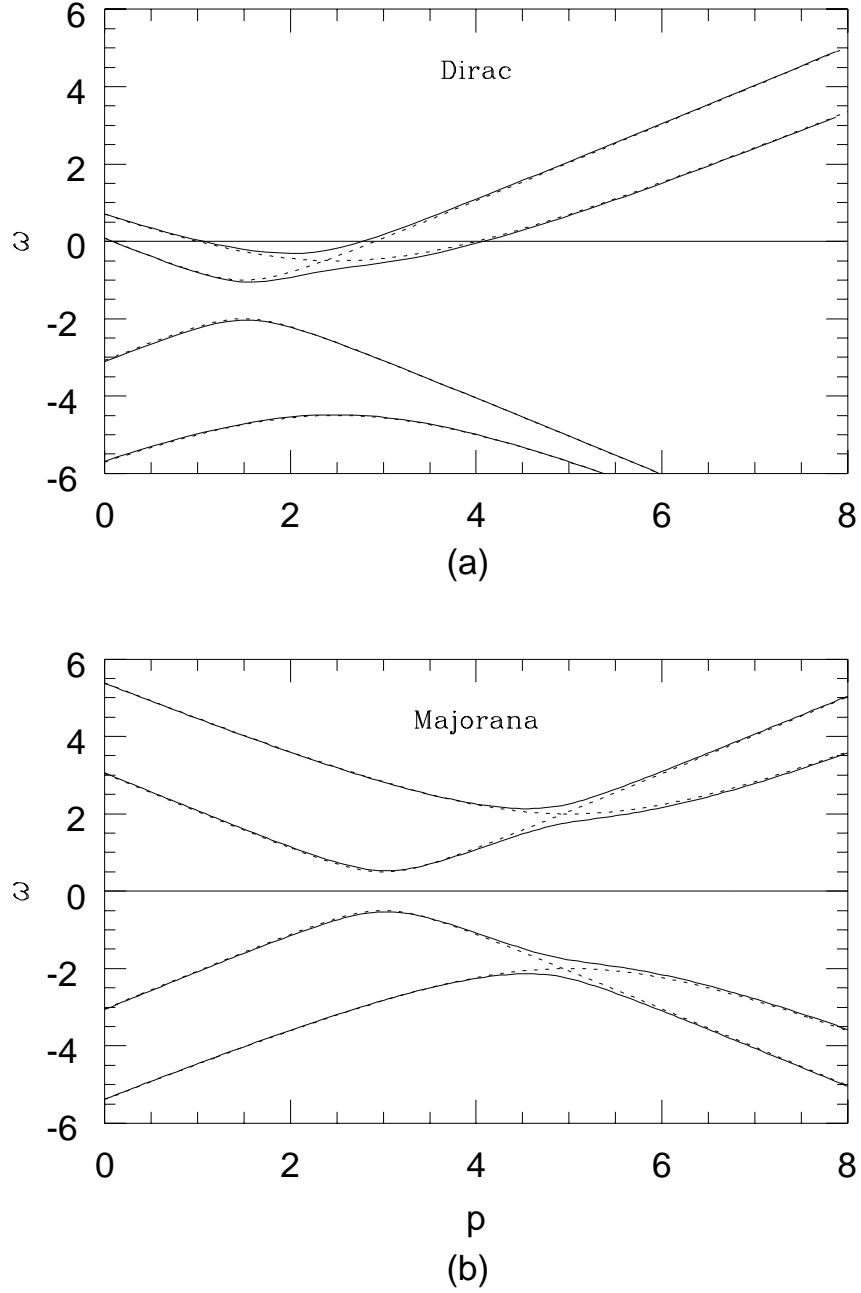}
\caption[Dispersion relations for two neutrinos]
{Dispersion relations for two neutrinos in (a) the Dirac case (negative
helicity) and (b) the Majorana case.  In both cases we have 
set $\alpha$$=$$1.0$, $\beta$$=$$2.5$, $m_1$$=$$0.5$ and 
$m_2$$=$$2.0$, in arbitrary units.  The dotted and solid curves 
correspond to $\theta$$=$$0$ and $0.2$, respectively.}
\label{fig:quartic}
\end{figure}

Fig.~\ref{fig:quartic}(b) shows a plot of the dispersion relations
for Majorana neutrinos in a medium.  
The dotted and solid curves correspond to the cases
with no coupling and with $\theta$$=$$0.2$, respectively.  Again
the curves with non-zero coupling are similar to those 
with no coupling, except for the ``level repulsion'' which 
occurs in the former case.
The parameters in this
plot are identical to those in the analogous plot
for Dirac neutrinos shown in Fig.~\ref{fig:quartic}(a).  Clearly
the dispersion relations are quite different in
the two cases.  

In all cases examined the minimum of the dispersion relations
is always greater than or equal to the minimum mass
and so again there appears to be no trapping in the Majorana case.

For relativistic neutrinos
the exact expressions for the energies may be simplified somewhat
to give
\begin{eqnarray}
        \omega & \simeq & p -(2\beta -\alpha)+\frac{\mu^2}{2p}
        \pm\frac{1}{4p}\left[\left(4\alpha
p-\Delta^2\cos(2\theta)\right)^2
        +\Delta^4\sin^2(2\theta)\right]^{1/2}, \label{exp1m}\\
        \omega & \simeq & -p +(2\beta -\alpha)-\frac{\mu^2}{2p}
        \mp\frac{1}{4p}\left[\left(4\alpha
p-\Delta^2\cos(2\theta)\right)^2
        +\Delta^4\sin^2(2\theta)\right]^{1/2}, \label{exp2m}
\end{eqnarray}
the first of which is in agreement with
the analogous expression, Eq.~(\ref{exp1}), for negative helicity
neutrinos in the Dirac case.

\subsection{The Vector Model}
\label{sec:vec}

\begin{figure}[tb]
\epsfbox[72 442 458 703]{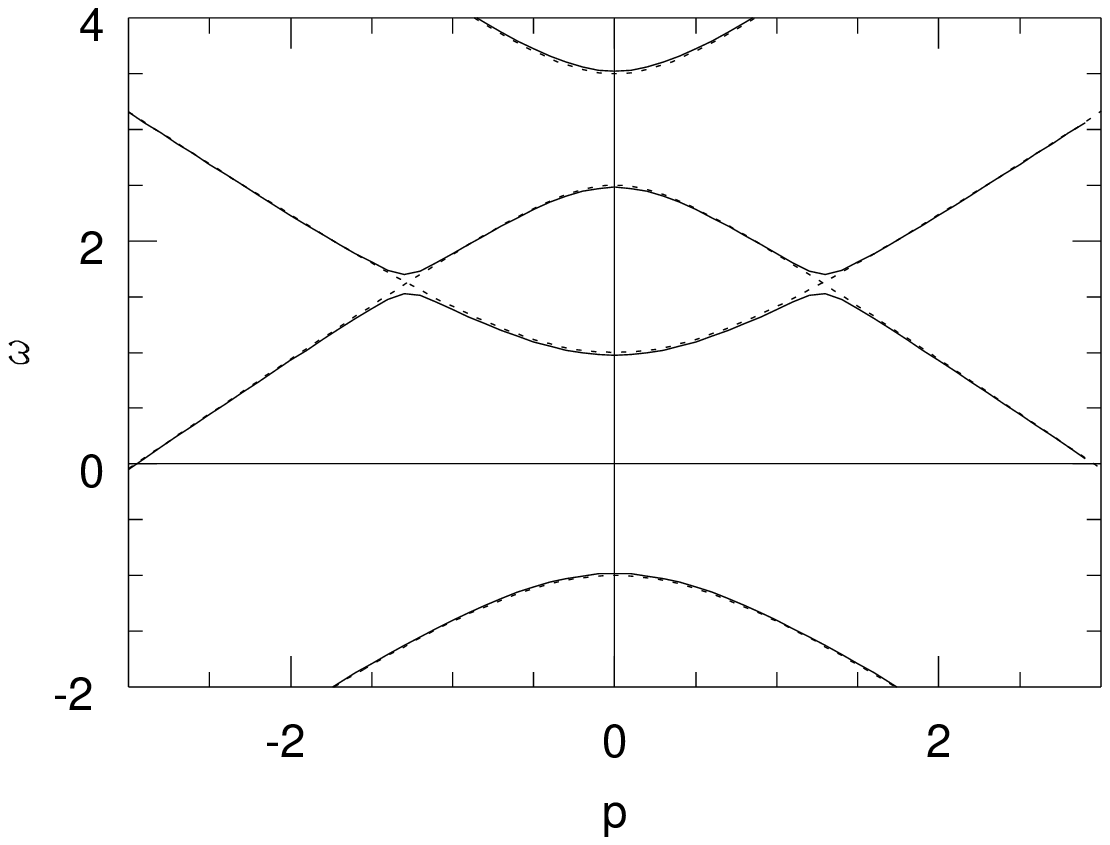}
\caption[Dispersion relations in the two-neutrino vector model]
{Dispersion relations in the two-neutrino vector model, with
$\alpha$$=$$3.0$, $m_1$$=$$0.5$ and
$m_2$$=$$1.0$, in arbitrary units.  The dotted and solid curves
correspond to $\theta$$=$$0$ and $0.2$, respectively.}
\label{fig:vector}
\end{figure}

In the models considered so far, the fact that the dispersion
relations have minima at non-zero values of the momentum
is due to the chiral nature of the potential in the effective
Dirac equation.  That is, since the potential depends on
the spin, the curves are displaced to the right or left 
depending on whether the neutrino's spin is parallel or 
anti-parallel to its momentum.  As we have seen, this phenomenon
occurs for a single neutrino flavour and persists when another
flavour is added.  It is amusing to note that it is possible
to obtain a minimum at non-zero momentum even with a purely
vector interaction, although this effect requires
the presence of at least two neutrino fields.  
Such a model was studied several years ago by
Chang and Zia~\cite{changzia}.  The effective Lagrangian is
in this case given by
\equation
        \left\{\pslash -M -\alpha Q \gamma^0\right\}\psi=0 ,
\endequation
where the matrix $Q$ is as defined in Eq.~(\ref{qdef}).  The
above equation is similar to Eq.~(\ref{deqn2neut})
except for the absence of the $(1-\gamma^5)$ factor which was present
in that case.  We have also set $\beta$$=$$0$ for simplicity.
The equations governing the dispersion relations
may be derived in a manner similar to the Dirac case above to yield
\begin{eqnarray}
	& &\left[\left(\omega-\alpha\cos^2\theta\right)^2-p^2-m_1^2\right]
	\left[\left(\omega-\alpha\sin^2\theta\right)^2
		-p^2-m_2^2\right]\nonumber \\
	& &\;\;\;\;=
	2\alpha^2\sin^2\theta\cos^2\theta\left(\omega^2+p^2-
		\alpha\omega+m_1m_2\right)+\alpha^4\sin^4\theta\cos^4\theta
\end{eqnarray}
which is independent of the spin and is symmetric under
$p$$\rightarrow$$-p$.
A-priori it might then seem impossible to generate a minimum at non-zero
$p$.  Indeed, for a single neutrino flavour this is the case.  For two
flavours, however, something very interesting can happen.
Suppose we first set $\theta$ to zero and imagine increasing
$\alpha$ by so much that the negative energy $\nu_e$ solution
overlaps with the positive energy $\nu_{\mu}$ solution.  When a 
non-zero coupling is included, these levels repel each other
and minima develop near the former crossing points, symmetrically
placed about the origin.  
This feature is illustrated in Fig.~\ref{fig:vector}.
Thus in this case as well it is
possible to have minima in the dispersion relations at non-zero
values of the momentum.  Note that this case is still somewhat
different from the chiral cases which we have studied above, since
the first and second derivatives at the origin are zero and
negative, respectively, corresponding to a negative effective
mass at the origin.  This is not the case in chiral theories.

\subsection{Neutrino Oscillations}
\label{sec:neutosc}

We have so far mostly restricted our attention to an investigation
of the forms of the dispersion relations themselves and have
not considered in detail the effects that these would have
on the oscillations of neutrinos.  We first note that in the
relativistic regime, the standard MSW results are recovered; that is,
(i)
the neutral current contribution factorizes and (ii) the
negative-helicity states obtain the appropriate dispersion
relations in matter while the positive-helicity states revert to their
vacuum dispersion relations.

For non-relativistic neutrinos, however, the situation is 
in some sense far more interesting.  One novel effect which arises
in the Dirac case purely as a result of the chiral nature of the  
potential is that in principle one could observe neutrino
oscillations {\em with only a single neutrino flavour}. 
This could happen since, for non-relativistic neutrinos, the 
left-handed interactions responsible for producing neutrinos
would produce both negative- and positive-helicity neutrinos.
Since these propagate with different phase velocities in the
medium, they would in general get out of phase
with each other, producing oscillations in the probability
to detect left-handed neutrinos.  For relativistic
neutrinos this effect disappears since the amplitude to
produce and detect a positive-helicity neutrino becomes negligible.
Note, however, that the difference in phase velocities
remains and would lead to oscillations if only positive-helicity 
neutrinos could be produced and detected.

The generalization of this effect to the two-neutrino
case gives the result that in general there could
be oscillations between four different states.
This would lead to an oscillation probability which is 
a superposition of four different oscillation curves.

\section{Discussion and Conclusions}
\label{sec:concl}

In this paper we have examined the coherent interactions
of a neutrino with a background medium by examining the 
solutions of the effective Dirac equation for the neutrino.
A close analysis revealed that the dispersion relations corresponding
to such a Dirac equation have a non-trivial form, even
in the simple case in which there is only a single neutrino
flavour.  In particular, we have examined the interesting effects
which arise due to the minimum of the dispersion relation which
occurs for non-zero momentum.  We have shown that, quite generally
for Dirac neutrinos,
the minimum value of the energy is less than the neutrino's mass,
which implies that for any such background there will be
trapping of very low energy neutrinos.   
In cases in which the strength of the
potential exceeds twice the rest mass of the neutrino
in vacuum, neutrino--antineutrino pairs are produced by the
electron density gradient at the boundary and this affects the
reflection amplitude of a low energy neutrino in the medium.
Our analysis of the case of a single Majorana neutrino flavour
showed that Majorana neutrinos are not trapped by the
medium.   

We have also presented a study of the dispersion
relations for Dirac and Majorana neutrinos for the Standard Model
with two neutrino flavours.  In this case we found that
the trapping phenomenon persists in the Dirac case but is absent
in the Majorana case.
The neutral current contribution to the oscillation
probabilities  ``factorizes'' in the relativistic 
regime, but not in the non-relativistic case.  We saw, in fact,
that in principle neutrino oscillations could occur
with only a {\em single} flavour of neutrino, due to the  
different phase velocities of the helicity eigenstates.

\acknowledgements{We wish to thank F. Goldhaber, H. Lipkin, S. Nussinov
and M. Tytgat for helpful conversations.  This work was supported in 
part by the Natural Sciences
and Engineering Research Council of Canada. Their support is gratefully
acknowledged. Ken Kiers is also grateful to the High Energy Theory Group
at Brookhaven National Laboratory
for support provided under contract number DE-AC02-76CH00016 with the
U.S. Department of Energy. Nathan Weiss acknowledges the support
of the Weizmann Institute and the Shrum Fund as well as the support of
this
work by the Israel Science Foundation under grant number 255/96-1 and
the US-Israel Binational Science Foundation under grant
number 94-314.}

\begin{thebibliography}{99}

\bibitem{msw} L. Wolfenstein, Phys. Rev. D {\bf 17} (1978) 2369;
	Phys. Rev. D {\bf 20} (1979) 2634; \\
	S.P. Mikheyev and A.Yu. Smirnov, Yad. Fiz. {\bf 42} (1985)
	1441 [Sov. J. Nucl. Phys. {\bf 42} (1985) 913]; Il Nuovo 
	Cimento C {\bf 9} (1986) 17.

\bibitem{mannheim} P.D. Mannheim, Phys. Rev. D{\bf 37} (1988) 1935.

\bibitem{nieves} J.F. Nieves, Phys. Rev. D{\bf 40} (1989) 866.

\bibitem{notraf} D. N{\"o}tzold and G. Raffelt, Nucl. Phys. {\bf B307}
	(1988) 924.

\bibitem{pantaleone} J. Pantaleone, Phys. Lett. B {\bf 268} (1991) 227;
        Phys. Rev. D {\bf 46} (1992) 510.

\bibitem{loeb} A. Loeb, Phys. Rev. Lett. {\bf 64} (1990) 115.

\bibitem{smirnov} Note that while the trapping effect which we discuss
	may be difficult to investigate directly via experiment, there
	could be less direct effects due to the trapped neutrinos
	which are nevertheless of interest.  There has, for example,
	been a recent discussion concerning the possible effect
	of trapped (Dirac) neutrinos on long-range interactions in
	stars:  A.Yu. Smirnov and F. Vissani, hep-ph/9604443;\\
	A.Yu. Smirnov, hep-ph/9611465;\\
	As. Abada, M.B. Gavela and O. P\`ene, Phys. Lett. B {\bf 387}
	(1996) 315;\\
	E. Fischbach, Ann. Phys. {\bf 247} (1996) 213.

\bibitem{thesis} K.A. Kiers, ``A study of neutrino propagation
	and oscillations both in vacuum and in dense media,''
	Ph.D. thesis at the University of British Columbia, 1996.

\bibitem{hansen} An interesting discussion on the interplay between
	particle reflection at an interface and pair production may
	be found in:  A. Hansen and F. Ravndal, Phys. Scripta {\bf 23} 
	(1981) 1036.

\bibitem{mann2} P.D. Mannheim, Int. J. Theor. Phys. {\bf 23} (1984) 643.
 
\bibitem{palpham} P.B. Pal and T.N. Pham, Phys. Rev. D {\bf 40} (1989)
259.

\bibitem{changzia} L.N. Chang and R.K.P. Zia, Phys. Rev. D {\bf 38}
        (1988) 1669.

\end{thebibliography}
\end{document}